\documentclass[12pt,pre,letterpaper,showpacs,superscriptaddress]{revtex4}
\def\be{\begin{equation}}
\def\fin{\end{equation}}
\def\disp{\displaystyle}

\def\T{{\sf T\kern-.45em T}}
\def\C{\kern.1em{\raise.47ex\hbox{$\scriptscriptstyle |$}}
             \kern-.40em{\sf C}}

\def\ze{\zeta}
\def\al{\alpha}
\def\b{\beta}

\def\hfl{\disp\mathop{\hbox to 10mm{\rightarrowfill}}}

\usepackage{graphicx,amssymb,amsmath}
\begin{document}
\title{Corrections to the Law of Mass Action and Properties of the Asymptotic $t = \infty$ State
for Reversible Diffusion-Limited Reactions}

\date{\today}

\author{R.Voituriez}
\email{voiturie@lptl.jussieu.fr}
\affiliation{Laboratoire de Physique Th{\'e}orique des Liquides,
Universit{\'e} Paris 6, 4 Place Jussieu, 75252 Paris, France}
\author{M.Moreau}
\email{moreau@lptl.jussieu.fr}
\affiliation{Laboratoire de Physique Th{\'e}orique des Liquides,
Universit{\'e} Paris 6, 4 Place Jussieu, 75252 Paris, France}
\author{G.Oshanin}
\email{oshanin@lptl.jussieu.fr}
\affiliation{Laboratoire de Physique Th{\'e}orique des Liquides,
Universit{\'e} Paris 6, 4 Place Jussieu, 75252 Paris, France}
\affiliation{
Max-Planck-Institut f\"ur Metallforschung, Heisenbergstr. 3,
D-70569 Stuttgart, Germany}
\affiliation{Institut f\"ur Theoretische und Angewandte Physik,
Universit\"at Stuttgart, Pfaffenwaldring 57, D-70569 Stuttgart,
Germany}

\begin{abstract}
On example of diffusion-limited reversible $A+A \rightleftharpoons
B$ reactions we re-examine two fundamental concepts of classical
chemical kinetics - the notion of "Chemical Equilibrium" and the
"Law of Mass Action". We consider a general model with
distance-dependent reaction rates, such that any pair of $A$
particles, performing standard random walks on sites of a
$d$-dimensional lattice and being at a distance $\mu$ apart of
each other at time moment $t$, may associate forming a $B$
particle at the rate $k_+(\mu)$. In turn, any randomly moving $B$
particle may spontaneously dissociate at the rate $k_-(\lambda)$
into a geminate pair of $A$s "born" at a distance $\lambda$ apart
of each other. Within a formally exact approach based on
Gardiner's Poisson representation method we show that
 the asymptotic $t = \infty$ state attained by
such diffusion-limited reactions is generally \textit{not a true
thermodynamic equilibrium}, but rather a non-equilibrium
steady-state, and that the Law of Mass Action is invalid. The
classical concepts hold \text{only} in case when the ratio
$k_+(\mu)/k_-(\mu)$ does not depend on $\mu$ for any $\mu$.
\end{abstract}


\maketitle

\section{Introduction}

"Chemical Equilibrium" (CE) and the "Law of Mass Action" (LMA) are
two central concepts of classical chemical kinetics (see, e.g.,
Refs.\cite{keizer,silbey}). In virtually every text-book one
finds, regarding, for instance, the behavior of reversible
association/dissociation reaction of the form \be\label{chem} A+A\
\disp \mathop{\rightleftharpoons}^{K_+}_{K_-} \ B, \fin where
$K_+$ and $K_-$ are the forward and the backward rate constants,
respectively, that the asymptotic state achieved in {\em closed }
systems at $t = \infty$ is the state of Chemical Equilibrium -
state with no net change in activity, or concentration with time
$t$. Thermodynamically, the condition of CE is the condition in
which the driving forces of the reaction in Eq.(\ref{chem}) (or
any other reversible reaction) are equal and opposite. This
condition implies that no spontaneous change is observed and that,
according to the zeroth principle of thermodynamics, the net Gibbs
free energy change of a
 mixture of reactants and products vanishes.
Kinetically,
the condition of CE is the condition
in which the rates of the forward, $K_+ a_{\infty}^2$, and the backward, $K_-
b_{\infty}$, reactions are equal and opposite, such that
$a_{\infty}$ and $b_{\infty}$ - the
 "equilibrium" concentrations of $A$ and $B$
species, obey the LMA:
\be
\label{lma}
-K_+ a_{\infty}^2 + K_- b_{\infty} = 0 \;\;\; \text{or} \;\;\;
\frac{a_{\infty}^2}{b_{\infty}} = \frac{K_-}{K_+} = K_{eq},
\fin
with $K_{eq}$ being the  "equilibrium" constant,
dependent only on the \textit{thermodynamic}
properties of the reactive system \cite{keizer,silbey}.
It is important to remark that once the CE is achieved,
the forward and backward reactions
continue to run. It is just at equilibrium,
since the rates are equal, there is no
visible or measurable change in the system.

In this paper we re-examine these two fundamental concepts,
serving to define the composition of reactive mixtures (as well as
general trends in reactive systems), in case of reversible
reactions which involve \textit{diffusive} particles and products,
i.e. the so-called \textit{diffusion-limited} reactions
\cite{fixman,deutch,bamberg,feld,lee,bursh}. Our focal questions
here are whether for such \textit{diffusion-limited} reactions
taking place in {\em closed} systems the CE is always a true thermodynamic equilibrium state (TES),
 and whether the LMA in
Eq.(\ref{lma}) always holds. We emphasize that it does not mean
that the principles of thermodynamics are contested: according to
the zeroth principle, for a closed system the CE always exists and
is by definition the thermodynamic equilibrium state. What is
meant here by \textit{true thermodynamic equilibrium} state, is
much more restrictive: within the conventional picture it
designates a state fully described by the thermodynamical, as
opposed to dynamical, quantities, and which satisfies the detailed
balance equilibrium. The fact that the CE is not, in some cases
described below, a TES is therefore fully compatible with the laws
of thermodynamics. Note that in particular the LMA is not a law of
thermodynamics, since it relies on an ideal gas approximation
which is not necessarily always the case.

We concentrate on a particular
reaction scheme - the simple association/dissociation reaction in
Eq.(\ref{chem}), but our analysis can be readily generalized for
any other type of reversible reaction. We consider here a rather
general lattice model of
 reactions in Eq.(\ref{chem}), which was first studied analytically by Zeldovich and Ovchinnikov [18].
 In this model the $A$ particles and products $B$
perform standard random walks on sites of a $d$-dimensional
hypercubic lattice and the elementary reaction rates are
long-ranged and dependent on the instantaneous distance between
any two $A$ particles; that is, any pair of $A$s may associate
(forming a $B$ particle) at any moment of time $t$ at rate
$k_+(\mu)$, where $\mu$ is the instantaneous distance separating
these two particles. In turn, any $B$ particle may spontaneously
dissociate at rate $k_-(\lambda)$ giving birth to a geminate pair
of $A$s separated by a distance $\lambda$. In our analysis we
suppose that the bimolecular elementary reaction rate $k_+(\mu)$
and the unimolecular elementary reaction rate $k_-(\lambda)$ are
arbitrary (integrable) functions of $\mu$ and $\lambda$. Note also
that we have chosen the lattice formulation just for the
convenience of exposition; an analogous continuous-space
formulation can be readily worked out. For this model, in terms of
a formally exact approach based on Gardiner's Poisson
representation method \cite{gardiner}, we obtain exact non-linear
Langevin equations describing the time evolution of complex-valued
Poisson fields, whose mean values determine the $A$ and $B$
particles' mean concentrations. Solutions of these Langevin
equations in the asymptotic $t = \infty$ state are obtained by two
different approaches: via a) a certain decoupling approximation
and b) a systematic diagrammatic expansion. From these solutions,
which coincide in the leading order, we deduce a general criterion
determining the conditions
 when the classical LMA in Eq.(\ref{lma}) holds and
when the asymptotic $t = \infty$ state is a true thermodynamic
equilibrium. We show that this may only happen when the
distance-dependent elementary reaction rates obey a rather strong
(and apparently unrealistic) condition: the ratio $
k_+(\mu)/k_-(\mu)$ does not depend on $\mu$ for any $\mu$! In case
when this microscopic restriction is violated even at a single
point, one can show that the detailed balance is broken.  Here we
demonstrate that this violation of the detailed balance at the
microscopic scale has macroscopic consequences: the LMA in
Eq.(\ref{lma}) is violated, particles' concentrations are
spatially correlated, the correlation length is macroscopically
large, and, remarkably, the CE is not a true thermodynamic
equilibrium but rather a non-equilibrium steady-state, depending
on dynamical properties such as the particles' diffusion
coefficient. The profound reason of these spectacular macroscopic effects is that the breaking of detailed balance generates a non--vanishing probability current which modifies and sustains the fluctuations. In turn these fluctuations correlate the particles' concentrations, perturbate the rate of the forward reaction and thus displace the equilibrium concentrations, breaking the LMA. Note that such a current occurs in the phase space and there is no net transport in the real space. We remark that this non-equilibrium steady-state
breaking the detailed balance provides an example of irreversible
circulation of fluctuations - a notion put forward in
Ref.\cite{tomita}. Using system size expansion method, Tomita and
Tomita \cite{tomita} have demonstrated that for statistical
physics systems a non-vanishing probability current can be
generated by breaking of the detailed balance (a state which they
call a "cyclic balance"). We
note parenthetically that appearance of a non-equilibrium steady
state is generic for open reaction-diffusion systems, as first
suggested by Kuramoto \cite{kuramoto} and subsequently elaborated
in Refs.\cite{nicolis,seki2,gorecki,wakou}. Under certain
conditions, it may also take place for reactions in which the
particles number is not explicitly conserved \cite{henkel}. We
emphasize that here such a non-equilibrium steady-state emerges in
a {\em closed} system with strictly conserved overall
concentration of particles and products, without any external
inflow of particles! This gives a striking example of a steady
state {\it breaking the detailed balance equilibrium},
characterized by non vanishing probability currents in the phase
space.

We note that despite the common belief that in closed systems the
CE is a true TES and that the LMA may be taken for granted, there
are some good reasons which might make our questions legitimate.
Indeed, the classical kinetic picture has already been proven to
be inadequate in many situations and
 a number of
significant deviations from the text-book behavior
has been discovered.
These deviations concern primarily the
 \textit{kinetic} behavior. For example, for reversible reactions
conventional chemical kinetics predicts an exponential
approach toward the CE state.
It has been realized, however,
that this is not the case for
reversible reactions involving diffusive species; here,
the concentrations approach
the asymptotic state $t \to \infty$
only
 as a power law ($t^{-d/2}$ in $d$ dimensions)!
Such a behavior stems out of many particle and non-linear effects.
Because of the bimolecular (non-linear) forward reaction, the time
evolution of observables - particles´ mean concentrations, appears
to be coupled to the evolution of the pairwise correlation
functions. In turn, long-time decay of correlations is dominated
by existence of the conserved values, e.g., the overall
concentration of particles and products, which are not affected by
reactions and thus represent purely diffusive modes of the system.
Theoretically, the power law approach to the asymptotic state has
been first predicted in Ref.\cite{ovch1} using physical arguments
based on the analysis of spatial concentration fluctuations, and
subsequently elucidated in terms of more elaborated approaches
\cite{ovch2,tachya,redner,gleb0,gleb,szabo,shin,naumann,gopich}.
In some cases, exact dynamical many-particle solutions have been
obtained \cite{cardy,gopich2,gopich3}. As well, this power law
behavior has
 been indeed observed
in excited state proton transfer reactions \cite{1,2},
and also seen in numerical MC simulations
\cite{3}.

The question whether in closed systems the LMA in Eq.(\ref{lma})
is valid for the reversible diffusion-limited reactions at $t =
\infty$ has already been raised in Refs.\cite{ovch2} and
\cite{gleb0,gleb}. In Refs.\cite{ovch2} an \textit{approximate}
approach has been proposed to describe kinetics of reversible
association/dissociation reactions in Eq.(\ref{chem}) with
distance-dependent elementary reaction rates. In this approach,
the evolution of the reactive system has been reduced to the
evolution of a mixture of two quantum Bose gases, which was solved
by extraction of  the condensate and approximate second
quantization method, similar to the Bogolyubov's theory of a
weakly non-ideal Bose gas \cite{bogol}. In Refs.\cite{ovch2}, some
 corrections to the LMA non-vanishing in the limit $t \to \infty$
have been found. Corrections to the LMA have also been obtained in
Ref.\cite{gleb0} in terms of a suitably extended Smoluchowski
approach for reversible, microscopically inhomogeneous contact
reactions, such that the reaction radius $R$ of the forward
reaction is unequal to the radius $\lambda$ of pairs appearing
within the course of the backward reaction. In three dimensions,
it was found that in the asymptotic $t = \infty$ state the
particles concentration obey \be \label{pp}
\frac{a_{\infty}^2}{b_{\infty}} = \frac{K_-}{K_+} \Big[ 1 +
\frac{K_+}{4 \pi D R} \left(1 - \frac{R}{\lambda}\right)\Big],
\fin where $D$ is particles diffusion coefficient. Note that
according to Eq.(\ref{pp}), particles concentrations in the
asymptotic $t = \infty$ state are dependent on such "kinetic"
parameter as $D$, which could be quite alarming, if Eq.(\ref{pp})
were derived in terms of a more reliable approach. Further on,
Refs.\cite{gleb} have pursued a different \textit{uncontrollable}
approach analyzing the temporal evolution of several contact
diffusion-limited reactions in terms of reaction-diffusion
equations for local concentrations. In these equations, the
stochastic nature of particles transport and reactions have been
incorporated by random source terms, derived
 within the framework of
a hydrodynamic-level  stochastic description of number densities for a dilute,
chemically reacting system Ref.\cite{grossmann}.
Solving the resulting non-linear Langevin equations
under an \textit{assumption} that the concentration fields are Gaussian (which implies
an automatic decoupling
of fourth-order correlations and hence, insures that the
third-order correlations vanish),
 Refs.\cite{gleb} predicted not only non-vanishing
corrections to the LMA, but also shown that in the asymptotic $t =
\infty$ state particles' spatial distributions are correlated.
Moreover, it has been realized that the correlation length depends
on particles' diffusion coefficients, which rules out, of course,
that the CE is a true TES, but rather represents a non-equilibrium
steady-state.

On the other hand, formally \textit{exact solutions} obtained
in Refs.\cite{cardy} and \cite{gopich3}
for several reversible diffusion-limited reactions
 have demonstrated that the LMA in Eq.(\ref{lma}) is strictly valid
at $t = \infty$, and that the asymptotic $t = \infty$ state
has a Poissonian spatial distribution of concentrations, which
signifies that the CE is a true TES. Consequently, approximate
\cite{ovch2,gleb0,gleb} and rigorous \cite{cardy,gopich3}
approaches, although agree on the dynamical behavior, are at some odds concerning
 the properties of the asymptotic $t =
\infty$ state.

Both exact approaches \cite{cardy,gopich3}, however,
focused on contact diffusion-limited reactions with some
rather restrictive conditions imposed
on the elementary reaction
acts. More specifically, in Ref.\cite{cardy},
which considered a continuous-space
model of different reversible reactions
and made use of suitably
generalized quantum field theory techniques
of Refs.\cite{droz,doi,peliti},
the forward reaction
radius and
the radius
of the geminate
pair born
in the backward reaction act were both set equal
to zero.
In Ref.\cite{gopich3}, which
used a lattice formulation ingeniously
extending
the Poisson representation method of Gardiner
\cite{gardiner}, it was stipulated
that the forward reaction takes
place when both species
occupy simultaneously the same
lattice site, while
the dissociation of the reaction product
produces a geminate pair of reactants born at
the \textit{same} lattice site. We note parenthetically that
for this particular case both Refs.\cite{ovch2} and
\cite{gleb0}, Eq.(\ref{pp}),
predict that corrections to the LMA vanish as $t \to \infty$ and thus
the LMA obeys its classical form in Eq.(\ref{lma}).

Strictly speaking, reaction events are not describable within the
framework of a classical theory only; an elementary reaction act
results from an interplay of many factors and is influenced by
solvent structure, potential interactions, a variety of particles'
energies and angular orientations, quantum processes of different
origin and etc \cite{hynes,moreau,hanggi,w,seki,moro}. At such
scales, a notion of a fixed "reaction radius" $R$ does not make
much sense - a "reactive" boundary condition, imposed at a fixed
distance $R$ separating the reactants, is just a mathematical
trick employed to obtain a tractable formalism and to circumvent
enormous technical difficulties. This technical breakthrough is
achieved, however, at expense of introducing a certain degree of
arbitrariness regarding the choice of the value of $R$ and of the
reaction rate itself, which both, in consequence, are rather
ill-defined. In this regard, for a more adequate description of an
elementary reaction act, (still being, however, in a reasonable
compromise between either too restrictive or too complicated
theoretical model) it is appropriate to introduce, as it has been
done, in particular, in Ref.\cite{ovch2}, a distance $\mu$
dependent elementary reaction rates for both forward and backward
reactions. Note that an account for the long-range
distance-dependent character of elementary reaction rates is
indispensable for the analysis of kinetics of reaction including
remote electron or proton transfer \cite{tach,burs}, for which the
contact approximation is meaningless. It appears, as well, that it
is indispensable also in the general case for the analysis of such
a delicate issue as the nature of the asymptotic $t = \infty$
state. The meaning of this statement will become clearer as we
proceed.

This paper is outlined as follows. In section 2 we formulate the
model,
write down the master equation describing
particles reactions and diffusion on the microscopic many particle level, and,
within the
framework of Gardiner's Poisson representation method \cite{gardiner},
derive non-linear Langevin equations describing the
time evolution of the Poisson fields.
In section 3
we discuss solutions of the non-linear Langevin
equations using a certain decoupling approximation. Within this approach, we
define a criterion
determining the conditions
 when the classical LMA in Eq.(\ref{lma}) holds and
when the asymptotic $t = \infty$ state is a true thermodynamic
equilibrium. Focusing next on a particular case of microscopically
inhomogeneous contact reactions, we determine explicitly
corrections to the LMA. Further on, in section 4, we set up a
systematic approximation scheme previously developed for contact
reactions in Ref.\cite{cardy}. We show, within this mathematically
rigorous approach, that the criterion obtained in section 3 is
exact. Moreover, we demonstrate that for microscopically
inhomogeneous contact reactions, the corrections to the LMA
obtained in section 3 within the decoupling approximation, as well
as the result in Eq.(\ref{pp}) obtained in Ref.\cite{gleb0} within
a suitably extended Smoluchowski approach, are \textit{exact} in
the linear order in deviation from the equilibrium. Besides, we
present here several explicit results for exponential reaction
rates.
 Finally, we conclude in section 5 with a brief summary of results and discussion.

\section{Model and Evolution Equations}

Consider an infinite $d$-dimensional hypercubic lattice of spacing
$\ell$, containing particles of two types - $A$ and $B$, which
perform unconstrained (an arbitrary number of particles can occupy
any lattice site) random walks between neighboring sites. Any two
$A$ particles, being at a distance $\mu$ from each other, may
enter into reaction at the rate $k_+(\mu)$ forming a single $B$
particle, placed at the half-distance between two $A$s. Further
on, any $B$ particle at any moment of time may spontaneously
dissociate at the rate $k_-(\lambda)$ producing a geminate pair of
$A$ particles born (with a random orientation) at a distance
$\lambda$ apart of each other. Note that the true reaction
constants, (those entering Eq.(\ref{lma})) are determined as
$K_{+} = \sum_{\mu} k_{+}(\mu)$ and $K_{-} = \sum_{\lambda}
k_{-}(\lambda)$ \cite{ovch2}. Most of our analysis will be
performed supposing that $k_{+}(\mu)$ and $k_{-}(\lambda)$ are
arbitrary integrable functions of $\mu$ and $\lambda$.

In what follows, we will distinguish between two
situations:\\
(i) the case of \textit{microscopically homogeneous} elementary
reactions, when $k_{+}(\mu)$ is strictly proportional to $
k_{-}(\mu)$ (or, in other words, when the ratio
$k_{+}(\mu)/k_{-}(\mu)$ is independent of $\mu$ for any $\mu$)\\
(ii) the general case of \textit{microscopically inhomogeneous}
reactions, where $k_{+}(\mu)/ k_{-}(\mu)$ is $\mu$-dependent, at
least in some region of space.

 The criterium of microscopically
homogeneous reactions has an important interpretation. Indeed, for
microscopically inhomogeneous reactions, any steady  distribution
of the concentrations of the species A and B breaks the detailed
balance equilibrium. This is due to the fact that in this case the
equilibrium constant $k_{-}(\mu)/ k_{+}(\mu)$ is space dependent :
equilibrium can not be satisfied at each point for homogeneous
concentrations. We will show below that this violation of the
detailed balance has important implications at the macroscopic
scale.

The state of the system at time $t$ is determined by
the time-dependent numbers $A(x)$  and $B(x)$
of $A$ and $B$ particles at site $x$ of the lattice.
The set of such numbers is denoted as
$\{A\}$ and $\{B\}$ and $P[A,B,t]$ stands for the probability of
finding the system at time moment $t$ in the $\{A\}$ and $\{B\}$ state.

Our analytical approach is based on the formally exact Poisson
representation method, proposed originally by Gardiner
\cite{gardiner}, and subsequently generalized in
Ref.\cite{gopich3} for the description of the fluctuation-induced
kinetics of reversible, microscopically homogeneous contact
diffusion-limited reactions. Extension of this approach to the
case of distance-dependent reactions is straightforward and here
we merely outline the steps involved.

The starting point of our analysis
is the following master equation describing the time evolution of
$P[A,B,t]$:
\begin{equation}
\label{ma}
\partial_t P[A,B;t]={\cal L}P[A,B;t],
\end{equation}
where ${\cal L}$ is an operator, accounting for
the reaction and diffusion processes
 on the microscopic,
many particle level. Explicitly, the right-hand-side of Eq.(\ref{ma}) of reads:
\begin{widetext}
\begin{eqnarray}
{\cal L}P[A,B;t] &=& \frac{D}{\ell^2}\sum_x\sum_{e_x}
    \Big[\Big(A(e_x)+1\Big) P[A(x)-1,A(e_x)+1,B,t] -  A(x)P[A,B,t] \Big]\nonumber \\
 &+& \frac{D}{\ell^2}\sum_x\sum_{e_x}
    \Big[\Big(B(e_x)+1\Big) P[A,B(x)-1,B(e_x)+1,t]-  B(x)P[A,B,t]\Big] - \nonumber \\
 &+&  \sum_\mu k_+(\mu)\sum_x
    \Big[ \Big(A(x-\frac{\mu}{2})+1+\delta_{\mu,0}\Big) \Big(A(x+\frac{\mu}{2})+1\Big) \times\nonumber\\
 &\times& P[A(x-\frac{\mu}{2})+1+\delta_{\mu,0},A(x+\frac{\mu}{2})+1+\delta_{\mu,0},B(x)-1,t]\nonumber\\
 &-&  A(x-\frac{\mu}{2})(A(x+\frac{\mu}{2})-\delta_{\mu,0})P[A,B,t] \Big] \nonumber\\
&+&  \sum_{\mu}
k_{-}(\mu)\sum_x \Big(B(x)+1\Big) P[A(x-\frac{\mu}{2})-1-\delta_{\mu,0},A(x+\frac{\mu}{2})-1-\delta_{\mu,0},B(x)+1,t]
\nonumber\\
&-& \sum_{\mu}
k_{-}(\mu)\sum_x B(x)P[A,B,t]
,
\label{ME2}
\end{eqnarray}
\end{widetext}
where $D = \ell^2/2 d \tau$ is $A$ and $B$ particles diffusion
coefficient, $\tau$ is a characteristic hopping time,  while  the
symbol $\sum_{e_x}$ denotes summation over all possible
orientations of the lattice vector  $e_x$.  As the initial
condition to Eq.(\ref{ME2}), we choose an uncorrelated Poisson
distribution on each site and for each species: \be
\label{initial} P[A,B,t=0] = e^{-A_0-B_0} \prod_{x}
\frac{A_0^{A(x)}}{A(x)!} \frac{B_0^{B(x)}}{B(x)!}, \fin where
$A_0$ and $B_0$ are mean occupation numbers of $A$ and $B$
particles per lattice site.

The next step consists in
projecting
the particles numbers on sites $x$
onto the Poisson states \cite{gardiner}:
\begin{eqnarray}
P[A,B,t] &=& \int \prod_x d \alpha(x) d\beta(x) \frac{\exp\left(-\alpha(x)\right) \alpha(x)^{A(x)}
}{A(x)!} \times \nonumber\\
&\times& \frac{\exp\left(-\beta(x)\right) \beta(x)^{B(x)}
}{B(x)!} F[\alpha(x),\beta(x),t]
\end{eqnarray}
Existence of such a transformation (and of the inverse one) is well established \cite{gardiner}.
As a matter of fact,
the factorial
moments of particles' numbers, say of $A(x)$,
are related to the Poisson fields $\alpha(x)$ and $\beta(x)$ via a simple formula:
\begin{eqnarray}\label{fact}
&&\left<A(x)\right>_n\equiv \left<A(x)\Big(A(x)-1\Big)...\Big(A(x)-(n-1)\Big)\right> = \nonumber\\
&=&\int   \disp\  \prod_{x} d\al(x) d\b(x)\ \al^n(x) F[\al(x),\b(x),t]=\left<\al^n(x)\right>,
\end{eqnarray}
where the angle brackets denote the probabilistic average
stemming out of stochastic reaction processes.
Note, however,
that the Poisson fields $\alpha(x)$ and $\beta(x)$ may be complex-valued;
consequently, $F[\alpha(x),\beta(x),t]$ may admit
negative values and hence can not be interpreted as a probability
distribution.

Defining next the generating function
\be
G[u(x),v(x),t]=\sum_{a,b}\prod_{x}u(x)^av(x)^b P[A,B,t]
\fin
one represents $G[u(x),v(x),t]$ as
\begin{eqnarray}
G[u(x),v(x),t] &=& \int   \disp\  \prod_{x} d\al(x) d\b(x)\
\exp[(u(x)-1)\al(x)] \times \nonumber\\
&\times& \exp[(v(x)-1)\b(x)]F[\al(x),\b(x),t],
\end{eqnarray}
which yields, eventually,
the following Fokker-Planck
equation determining evolution of $F[\alpha(x),\beta(x),t]$:
\begin{eqnarray}\label{FP1}
\partial_t F[\al(x),\b(x),t] &=&
- \disp \sum_{x} \sum_{\mu}
\Big\{\frac{\partial}{\partial\al(x)} C_1[\al,\b]+\frac{\partial}{\partial\b(x)} C_2[\al,\b]
- \nonumber\\
&-&\disp \frac{1}{2} \frac{\partial^2}{\partial\al(x-\mu/2)\partial\al(x+\mu/2)}
C_3[\al,\b] \Big\} F[\al(x),\b(x),t]
\end{eqnarray}
where
\be
\left\{\begin{array}{l}
\label{A}
C_1[\al,\b]=-\disp
k_{+}(\mu)\al(x)\Big(\al(x-\mu)+\al(x+\mu)\Big) +
k_{-}(\mu)\Big(\b(x-\mu)+\b(x+\mu)\Big)+D\Delta\al,\\
C_2[\al,\b]=\disp
k_{+}(\mu)\al(x-\mu/2)\al(x+\mu/2) -
k_{-}(\mu)\b(x)+D\Delta\b,\\
C_3[\al,\b]=\disp
k_{+}(\mu)\al(x+\mu/2)\al(x-\mu/2)-k_{-}(\mu)\b(x),
\end{array}
\right. \fin where $\Delta$ is the second finite difference
operator.

Note that Eq.(\ref{FP1}) can not be, of course, solved exactly. On
the other hand, we are in position to draw one important
conclusion directly from the form of Eq.(\ref{FP1}). By
inspection, one may verify that because of the presence of "mixed"
derivatives $\partial^2/\partial\al(x-\mu/2)\partial\al(x+\mu/2)
...$, in the microscopically inhomogeneous case, when
$k_{+}(\mu)/k_{-}(\mu)$ is $\mu$-dependent, this Fokker-Planck
equation does not admit  any stationary solution of the form: \be
F[\al(x),\b(x),t]= \prod_{x}
\delta[\al(x)-\left<\al\right>]\delta[\b(x)-\left<\b\right>] \fin
Therefore, if $k_{+}(\mu)/ k_{-}(\mu)$ depends on $\mu$, the
stationary distribution $P[A,B,t]$ is \textit{not} a product of
uncorrelated Poisson distributions, as one obtains in case of
contact, microscopically homogeneous reactions. This implies, in
turn, that the detailed balanced is not satisfied, which hints us
that here the asymptotic $t = \infty$ state might be rather
particular. Indeed, below we will demonstrate that the fact that
$F[\al(x),\b(x),t]$ does not factorize into the product $\prod_x
f[\al(x),\b(x),t]$, where each multiplier $f[\al(x),\b(x),t]$
depends only on the Poisson fields $\al(x)$ and $\b(x)$ on the
site $x$, implies that the CE state is not a true equilibrium
state and that the pairwise correlations in particles' spatial
distribution exist, which, in consequence, results in violation of
the LMA in Eq.(\ref{lma}).

Using It\^o's equivalence, we find the following  Langevin
equations corresponding to equation (\ref{FP1}):
\be\label{lan}
\left\{\begin{array}{l}
\disp \partial_t \al(x)=  \disp\sum_{\mu} C_1[\al,\b] + \ze(x,t),\\
\disp \partial_t \b(x)=  \disp\sum_{\mu} C_2[\al,\b],
\end{array}
\right. \fin where $\ze(x,t)$ is a Gaussian noise with zero mean
value, whose correlation is given by
\begin{eqnarray} \label{noises}
\Big<\ze(x,t)\ze(x+x',t')\Big>&=&\disp \delta(t-t')\Big[
k_{-}(x')\Big<\b(x')\Big> \nonumber\\
&-&k_{+}(x')\Big<\al(x-x'/2)\al(x+x'/2)\Big>
\Big].
\end{eqnarray}
Before we proceed further, some comments are in order. We first
remark that exactly the same Langevin equations can be obtained
extending to the general case with distance-dependent elementary
reaction rates the field theoretical technique, developed in
Ref.\cite{cardy} for contact microscopically homogeneous
reactions. We believe, however, that the approach used here is
more simple and transparent. Second, we note that contrary to the
situation with contact microscopically homogeneous reactions (see
Eqs.(12) and (13) in Ref.\cite{cardy}), in the general case when
the ratio $k_{+}(\mu)/k_{-}(\mu)$ is $\mu$-dependent, the
amplitude of the noise correlation can not be expressed as a time
derivative of $<\al(x)>$; hence, noise correlation does not vanish
as $t \to \infty$. Finally, we note that non-linear Langevin
equations have been already used in Refs.\cite{gleb0,gleb} to
analyze the temporal evolution of several contact
diffusion-limited reactions. Noise terms, employed in this
approach, were previously derived within the hydrodynamic-level
stochastic approach in Ref.\cite{grossmann}. Our Eqs.(\ref{lan})
and (\ref{noises}) have essentially the same structure
as those used in Refs.\cite{gleb0,gleb} with two notable exceptions:\\
(i) Eqs.(\ref{lan}) and (\ref{noises}) are obeyed by the Poisson
fields, which are not physical concentrations, and moreover, the noise is
complex-valued, while the corresponding Langevin equations in
Refs.\cite{gleb0,gleb} are formulated for the true local
concentrations.\\ (ii) the noise terms used in
Refs.\cite{gleb0,gleb} contain an additional, compared to our
Eq.(\ref{noises}), term $\sim D \nabla_x A(x) \nabla_x \delta(x)$,
which does not vanish as $t \to \infty$.  This is apparently
incorrect and represents an evident shortcoming of  the
hydrodynamic-level stochastic approach in Ref.\cite{grossmann}.
Moreover, this is precisely the reason why the deviations from the LMA were
obtained in Refs.\cite{gleb0,gleb} for \textit{ microscopically homogeneous}
contact reactions.

\section{The asymptotic $t = \infty$ state: Decoupling approximation.}

For convenience, we turn  from now on to the continuous-space limit ($\ell \to 0$),
consider $x$ as a continuous variable
 and operate with particles'
concentrations
$a(x) = A(x)/l^d$ and $b(x) = B(x)/l^d$, instead of particles'
numbers. Note also that for an accurate "passage" to such a limit,
one has to introduce the rescaled reaction rates $k_{+}\to k_{+}\ell^{d-1}$ and $k_{-}\to k_{-}\ell^{-1}$, noise $\ze\to\ze/l^d$, as well as the Poisson
fields $\al(x)\to\al(x)/l^d$ and $\b(x)\to\b(x)/l^d$ \cite{cardy}.

Taking into account that,
in virtue of Eq.(\ref{fact}), we have
$< \alpha(x) > = <a(x)> = a$ and  $< \beta(x) > = <b(x)> = b$,
we write the Poisson fields as follows:
\begin{equation} \label{dev}
\begin{array}{l}
\alpha(x,t)=a+\delta\al(x,t),\\
\b(x,t)=b+\delta\b(x,t)
\end{array}\fin
where $a$ and $b$ are (time-dependent) particles' mean concentrations, while
$\delta\al(x,t)$ and $\delta\b(x,t)$ denote local deviations
of the Poisson fields from their mean values. Hence, by definition,
$< \delta\alpha(x,t)> = <\delta\beta(x,t)> = 0$.

Substituting next expressions in Eq.(\ref{dev}) into Langevin Eqs.(\ref{lan}), averaging
the resulting
equations and using the definition of the overall reaction constants
$K_{\pm} = \int_{\mu} k_{\pm}(\mu) \, d\mu$, we find that
particles' mean concentrations obey:
\begin{eqnarray} \label{exa}
\partial_t a &=& - 2 K_+ a^2 + 2 K_- b + 2 \Omega(0), \nonumber\\
\partial_t b &=&  K_+ a^2 - K_- b - \Omega(0),
\end{eqnarray}
where
\be \label{Omega}
\Omega(0) \equiv - \disp \int d\mu k_{+}(\mu)\sigma_{\alpha \alpha}(\mu),
\fin
and $\sigma_{\alpha \alpha}(\mu)$ is the pairwise correlation function,
\be\label{cor}
\sigma_{\alpha \alpha}(\mu) = \Big< \delta\alpha(x-\mu/2,t) \delta\alpha(x + \mu/2,t) \Big>,
\fin
of the fluctuations in the Poisson fields.  The
corresponding correlation function for the fluctuations
in particles concentrations in the asymptotic $t = \infty$ state can be readily expressed
in terms of $\sigma_{\alpha \alpha}(\mu)$:
\be c(\mu)_{\infty}=\Big<\Big(a_{\infty} - a(x,t=\infty)\Big) \Big(a_{\infty} - a(x + \mu,t=\infty)\Big)\Big> = \sigma_{\alpha \alpha}(\mu)_{\infty}+a_\infty
\delta(\mu) \fin

Note that Eqs.(\ref{exa}) are formally exact for any $t$.
These equations show
that the time evolution of observables - the particles' mean
concentrations, is ostensibly coupled
to the evolution of pairwise correlations in the
reaction-diffusion system under study.

Turning next to the infinite time limit, $t \to \infty$, we get, slightly rearranging Eqs.(\ref{exa}), that
the  particles' mean concentrations
obey:
\be\label{eps}
\disp \frac{a_{\infty}^2}{b_{\infty}} = \frac{K_{-}}{K_{+}} + \frac{\Omega(0)_{\infty}}{K_{+}
b_{\infty}},
\fin
where the subscript $\infty$ signifies that we
deal with the asymptotic $t = \infty$ solution.

Note that Eq.(\ref{eps}) resembles the classical
Law of Mass Action
in Eq.(\ref{lma}), but differs from it due
to an extra term ${\Omega(0)}/K_{+} b_{\infty}$, which is dependent on the
pairwise correlation function $\sigma_{\alpha \alpha}(\mu)_\infty$ and embodies all
non-trivial physics associated with the fluctuation effects.
Therefore, one may judge directly from Eq.(\ref{eps}) that
the classical LMA holds only if $A$ particles distribution is spatially
uncorrelated
in the asymptotic $t = \infty$ state, i.e.
$\sigma_{\al \al}(\mu)_\infty \equiv 0$.

Now, our aim is to determine the pair correlation
function $\sigma_{\alpha \alpha}(\mu)$.
This turns out to be, however, quite a complicated problem. More
specifically,
when one writes the evolution equations
obeyed by pairwise correlations, he gets that these are coupled
to the third-order correlations
and etc, which is a signature of a genuine many
particle problem.

In this section we resort to an approximate approach, while some
exact results will be presented in the next section. Our approach
here is based on the assumption that the deviations
$\delta\al(x,t)$ and $\delta\b(x,t)$ are Gaussian random fields,
which implies an automatic decoupling of fourth-order correlations
into the product of pairwise ones, and hence, truncation of the
hierarchy of coupled reaction-diffusion equations on the level of
third-order correlations. Such an approach has been first employed
in Ref.\cite{bur} to  obtain the $t^{-d/4}$-law describing the
fluctuation-induced
 kinetics of irreversible $A + B \to 0$
reactions, and subsequently generalized in Refs.\cite{gleb0} and \cite{gleb} for
other irreversible and reversible
reactions.

Within this approach, we find that pairwise
correlations obey the following
system of equations:
\begin{eqnarray}
\label{j}
\partial_t \sigma_{\alpha \alpha}(\mu) &=& 2 D \Delta_{\mu} \sigma_{\alpha \alpha}(\mu) - 2 a
\int d\lambda \, k_{+}(\lambda) \Big( 2 \sigma_{\alpha \alpha}(\mu) +
\sigma_{\alpha \alpha}(\mu - \lambda) +
\sigma_{\alpha \alpha}(\mu + \lambda) \Big) \nonumber\\
&+& 2 \int d\lambda \, k_{-}(\lambda) \Big(\sigma_{\alpha \beta}(\mu-\lambda)
+ \sigma_{\alpha \beta}(\mu + \lambda)\Big) + \nonumber\\
&+& k_{-}(\mu) b -  k_{+}(\mu) a^2 - k_{+}(\mu)
\sigma_{\alpha \alpha}(\mu), \\
\partial_t \sigma_{\alpha \beta}(\mu)&=& 2 D \Delta_{\mu} \sigma_{\alpha \beta}(\mu) -
 \Big(2 a K_{+} + K_{-}\Big) \sigma_{\alpha \beta}(\mu)
- a \int d\lambda \, k_{+}(\lambda)
\Big(\sigma_{\alpha \beta}(\mu + \lambda) + \sigma_{\alpha \beta}(\mu - \lambda)\Big)
\nonumber\\
&+& a \int d\lambda \, k_{+}(\lambda) \Big(\sigma_{\alpha \alpha}(\mu + \frac{\lambda}{2}) +
\sigma_{\alpha \alpha}(\mu - \frac{\lambda}{2})\Big) + \nonumber\\
&+& \int d\lambda \, k_{-}(\lambda) \Big(\sigma_{\beta \beta}(\mu + \lambda) +
\sigma_{\beta \beta}(\mu - \lambda)\Big),
\end{eqnarray}
and
\begin{eqnarray}
\label{jj}
\partial_t \sigma_{\beta \beta}(\mu) = 2 D \Delta_{\mu} \sigma_{\beta \beta}(\mu)
- 2 K_{-}
\sigma_{\beta \beta}(\mu)
+
2 a \int d\lambda \,
k_{+}(\lambda) \Big(\sigma_{\alpha \beta}(\mu + \frac{\lambda}{2}) +
\sigma_{\alpha \beta}(\mu - \frac{\lambda}{2})\Big),
\end{eqnarray}
where the $"\alpha\beta"$ and $"\beta\beta"$ pair correlation functions are defined
as
\be
\sigma_{\alpha \beta}(\mu) = \Big<\delta \alpha(x-\mu/2,t) \delta \beta(x+\mu/2,t)\Big>,
\fin
and
\be
\sigma_{\beta \beta}(\mu) = \Big<\delta \beta(x-\mu/2,t) \delta \beta(x+\mu/2,t)\Big>.
\fin
Further on, introducing a pair of Fourier transforms:
\be
f(p) = \int d\mu \; e^{i (p \cdot \mu)} f(\mu), \;\;\; \text{and} \;\;\;
f(\mu) = \frac{1}{(2 \pi)^d} \int dp \; e^{- i (p \cdot \mu)} f(p), \fin
we get from Eqs.(\ref{j}) to (\ref{jj}) that $\sigma_{k l}(p)$, ($k,l = \alpha,\beta$),
 obey:
\begin{eqnarray}
\label{j1}
\partial_t \sigma_{\alpha \alpha}(p) &=& - 2 \Big( D p^2  + 2 a K_{+} +
2 a k_{+}(p)\Big)
\sigma_{\alpha \alpha}(p) \nonumber\\
&+& 4 k_{-}(p) \sigma_{\alpha \beta}(p)
+  k_{-}(p) b - k_{+}(p) a^2 +  \Omega(p), \\
\partial_t \sigma_{\alpha \beta}(p)&=& - \Big(2 D p^2
+ 2 a K_{+} + K_{-} +2 a k_{+}(p)\Big)
\sigma_{\alpha \beta}(p) \nonumber\\
&+& 2 a k_{+}(p/2) \sigma_{\alpha \alpha}(p) +
2 k_{-}(p) \sigma_{\beta \beta}(p),\\
\label{j3}
\partial_t \sigma_{\beta \beta}(p) &=&
4 a k_{+}(p/2) \sigma_{\alpha \beta}(p) - 2 \Big(D p^2 + K_{-}\Big) \sigma_{\beta \beta}(p),
\end{eqnarray}
where we have assumed that $k_{+}(\mu)$ and $k_{-}(\mu)$
depend only on the absolute value of $\mu$, and denoted
\be
\label{om}
\Omega(p) = - \int d\mu \, e^{i (p \cdot \mu)} k_{+}(\mu) \sigma_{\alpha \alpha}(\mu).
\fin
Turning now to the asymptotic limit $t = \infty$, we get from Eqs.(\ref{j1}) to (\ref{j3})
that $\sigma_{\alpha \alpha}(p)_{\infty}$ is determined as the solution of the following equation:
\begin{eqnarray}
\label{imp}
\sigma_{\alpha \alpha}(p)_{\infty} = F(p)  \Big[\Omega(p)_\infty-\epsilon(p)_\infty\Big],
\end{eqnarray}
where
\be\label{eff}
F(p) = \frac{X_2(p) \left( X_1(p) + X_2(p)\right) - X_3(p)}{2\left(
X_1(p) X_2(p) - X_3(p)\right) \left(X_1(p) + X_2(p)\right)},
\fin
\begin{eqnarray}
X_1(p) &=& D p^2 + 2 K_+ a_{\infty} + 2 k_+(p) a_{\infty}, \nonumber\\
X_2(p) &=& D p^2 + K_{-},\nonumber\\
X_3(p) &=& 4 k_{+}(p/2) k_{-}(p) a_{\infty},
\end{eqnarray}
and
\be\label{epsilon}
\epsilon(p)_{\infty} =k_{+}(p) a^2_{\infty}- k_{-}(p) b_{\infty}.
\fin
Note that Eq.(\ref{imp}) only \textit{implicitly} defines $\sigma_{\alpha \alpha}(p)_{\infty}$,
since the term $\Omega(p)_{\infty}$ on the right-hand-side of this equation
is itself dependent on
$\sigma_{\alpha \alpha}(p)_{\infty}$, Eq.(\ref{om}). Therefore, Eq.(\ref{imp}) is an integral equation
and its solution may be found only if we specify $k_{+}(\mu)$ and $k_{-}(\mu)$.

\subsection{Criterion of validity of the LMA and of the True Thermodynamic Equilibrium: Decoupling Approximation.}

Despite the fact that $\sigma_{\alpha \alpha}(p)_{\infty}$ has been so far only implicitly defined,
and in order to get some explicit results we have to fix $k_{+}(\mu)$ and $k_{-}(\mu)$,
we are
in position now to deduce a general criterion specifying
when the LMA in Eq.(\ref{lma}) holds and when it is violated.
To do this, we have merely to define conditions when $\sigma_{\alpha \alpha}(p)_{\infty} \neq 0$.

We proceed as follows. First, we set in Eq.(\ref{imp})
$\sigma_{\alpha \alpha}(p)_{\infty} \equiv 0$ (i.e. $\sigma_{\alpha \alpha}(\mu)_{\infty} \equiv 0$).
This implies that $\Omega(p)_{\infty} \equiv 0$, and hence, in order to have
$\sigma_{\alpha \alpha}(p)_{\infty} \equiv 0$, the following relation should hold as an identity:
\be \label{cond1}
k_{-}(p) b_{\infty} \equiv k_{+}(p) a^2_{\infty} \;\;\; \text{for any $p$},
\fin
or, in the $\mu$-domain,
\be \label{cond2}
k_{-}(\mu) b_{\infty} \equiv k_{+}(\mu) a^2_{\infty} \;\;\; \text{for any $\mu$.}
\fin
Note that in this only case the amplitude of the
noise-noise correlation in Eq.(\ref{noises}) is proportional to $\partial_t a$ and hence, vanishes as $t
\to \infty$.

Now, one readily verifies that the identities in Eqs.(\ref{cond1})
and (\ref{cond2}) may hold only if the reactions rates are identic
functions of $\mu$, i.e. the elementary reactions are
microscopically homogeneous. If this condition is violated, even
at least at a single point, pair correlations $\sigma_{\alpha
\alpha}(\mu)_{\infty}$ are non-zero, and hence, in virtue of
Eq.(\ref{eps}), the LMA in Eq.(\ref{lma}) is violated. Hence, we
find that the conventional LMA in Eq.(\ref{lma}) is always
violated for microscopically inhomogeneous reactions.

Next, for microscopically inhomogeneous reactions, we estimate the
decay of pairwise correlations in the limit $\mu \to \infty$. In
the limit $p \to 0$, ($\mu \to \infty$), the kernel $F(p)$ in
Eq.(\ref{eff}) attains the form: \be F(p) \sim \disp \frac{2 D^2
p^4 + D p^2 \Big(4 K_{+} a_{\infty} + 3 K_{-}\Big) + K_{-}^2}{2 D
p^2 \Big(D p^2 + 4 K_{+} a_{\infty} + K_{-}\Big)\Big( 2 D p^2 + 4
K_{+} a_{\infty} + K_{-}\Big)} \fin Now, let $\Lambda_{r}$ be the
characteristic decay length of $k_{+}(\mu)$. Since this property
is usually of a range of a few interparticle separations, one may
expect that $\sigma_{\alpha \alpha}(\mu)_{\infty}$ varies with
$\mu$ much more slowly than $k_{+}(\mu)$. This assumption will be
checked for consistency afterwards. Hence, we may approximate
$\Omega(p)_{\infty}$ in Eq.(\ref{om}) as \be \label{40}
\Omega(p)_{\infty} \sim - \sigma_{\alpha \alpha}(\mu = 0)_{\infty}
k_{+}(p) \fin Substituting the latter representation into
Eq.(\ref{imp}) and performing straightforward calculations, we get
that in the limit $\mu \to \infty$, the pairwise correlations in
the asymptotic $t = \infty$ state obey \be \label{cl}
\sigma_{\alpha \alpha}(\mu)_{\infty} \sim \exp\Big(-
\frac{\mu}{\Lambda_{corr}}\Big)/\mu, \fin where \be \label{cl1}
\Lambda_{corr} = \sqrt{\frac{D}{4 K_{+} a_{\infty} + K_{-}}}. \fin
The approximation underlying the derivation of Eqs.(\ref{cl}) and
(\ref{cl1}) is thus justified when $\Lambda_{corr} \gg
\Lambda_{r}$, or, in other words, when $\tau_{chem} = (4 K_{+}
a_{\infty} + K_{-})^{-1} \gg \tau_{diff} = \Lambda_{r}^2/D$.

Hence, pairwise correlations in $A$ particles distributions are long-ranged for microscopically
 inhomogeneous reactions. A salient feature of the result in Eq.(\ref{cl}) is that the
  characteristic decay length $\Lambda_{corr}$ of pairwise correlations depends on such
   "kinetic" parameter as $D$, the particles´ diffusion coefficient, which signifies that the
   asymptotic state achieved  by reversible diffusion-limited reactions with
    microscopically inhomogeneous elementary reaction rates at $t = \infty$ is not a true
    equilibrium state, but rather a non-equilibrium steady-state.
We note finally that $\Lambda_{corr}$ is, of course, $D$-dependent
for arbitrary relation
 between $\tau_{chem}$ and $\tau_{diff}$. Note also that for
microscopically inhomogeneous reactions $\Lambda_{corr}$ does not
vanish as $D \to 0$; in this case, the particles may perform
random excursions in space just because in the break-up of $B$ a
pair of $A$s is not forced to be born at their initial locations,
at which they have entered the reaction forming the $B$ particle.
It is also interesting to remark  that in the limit $\tau_{chem} =
(4 K_{+} a_{\infty} + K_{-})^{-1} \gg \tau_{diff} =
\Lambda_{r}^2/D$ the correlation length $\Lambda_{corr}$ equals
the distance travelled by a diffusive particle within its typical
life-time between the reaction processes, i.e. the so-called
Kuramoto length. Remarkably, precisely this length sets the scale
of spatial correlations in out of equilibrium open chemical
systems \cite{kuramoto,nicolis,seki2,gorecki,wakou}.

\subsection{"Contact reaction" approximation for microscopically inhomogeneous reactions.}

In this subsection we consider,
within the framework of the decoupling procedure,
the properties of the asymptotic $t \to \infty$ state for
reversible diffusion-limited reactions in "contact approximation".
This will allow us to illustrate the statement made in the Introduction
that this approximation entails a somewhat ambiguous definition of the
properties of the asymptotic $t = \infty$ state, as well as to present
some explicit results on the
corrections
to the LMA in Eq.(\ref{lma}).

Suppose that $k_+(\mu)$ and $k_-(\mu)$ are some bell-shaped functions
centered around
their most probable values $R$ and $\lambda$, and
the thicknesses of the distributions
are negligibly small, such that
the forward and the backward distance-dependent reaction constants can be deemed
as delta-functions:
$k_+(\mu)= \gamma_d^{-1}(R) K_{+} \delta^d(|\mu| - R)$ and
$k_-(\mu)=\gamma_d^{-1}(\lambda) K_- \delta^d(|\mu|-\lambda)$, where the normalization
factor $\gamma_d(R) = A_d R^{d-1}$, $A_d$ being
the volume of a $d$-dimensional unit sphere, $A_d
= 2 \pi^{d/2}/\Gamma(d/2)$, and $\Gamma(x)$ - the Gamma-function.

For such a choice of the reaction rates, we get from
Eq.(\ref{imp}) that \be \sigma_{\alpha \alpha}(p)_{\infty} = F(p)
\Big(k_{-}(p) b_{\infty} - k_{+}(p) a_{\infty}^2 - k_{+}(p)
\sigma_{\alpha \alpha}(|\mu| = R)_{\infty} \Big), \fin which
implies \be \label{prop} \sigma_{\alpha \alpha}(\mu)_{\infty} =
J_{-}(\mu,\lambda) b_{\infty} - J_{+}(\mu,R) a_{\infty}^2 -
J_{+}(\mu,R) \sigma_{\alpha \alpha}(|\mu| = R)_{\infty}, \fin
where \be J_{-}(\mu,\lambda) = \frac{1}{(2 \pi)^d} \int dp \; e^{-
i (p \cdot \mu)} F(p) k_{-}(p), \fin and \be J_{+}(\mu,R) =
\frac{1}{(2 \pi)^d} \int dp \; e^{- i (p \cdot \mu)} F(p)
k_{+}(p). \fin Next, setting $|\mu| = R$ in Eq.(\ref{prop}), we
obtain the following equation determining the value of the
pairwise correlation function $\sigma_{\alpha \alpha}(\mu)$ at
distance equal to the direct reaction radius: \be \label{prop1}
\sigma_{\alpha \alpha}(|\mu| = R)_{\infty} = \disp
\frac{J_{-}(|\mu| = R,\lambda) b_{\infty} - J_{+}(|\mu|=R,R)
a_{\infty}^2}{1 + J_{+}(|\mu|=R,R)} \fin Consequently, taking
advantage of Eqs.(\ref{eps}) and (\ref{prop1}), we find that for
reactions in contact approximation particles'  mean concentrations
$b_{\infty}$ and $a_{\infty}$ obey: \be \label{act} \disp
\frac{a_{\infty}^2}{b_{\infty}} = \frac{K_{-}}{K_{+}}\left(1 +
J_{+}(|\mu|=R,R) - \frac{K_{+}}{K_{-}}
J_{-}(|\mu|=R,\lambda)\right). \fin Note now that setting $R
\equiv \lambda$,  we get that $J_{+}(|\mu|=R,R) = K_{+} J(R)$ and
$J_{-}(|\mu|=R,\lambda=R) = K_{-} J(R)$, which immediately implies
that Eq.(\ref{act}) simplifies to the conventional LMA in
Eq.(\ref{lma}) and signifies that the chemical equilibrium state
is a true thermodynamic equilibrium. If we, however, take into
account dispersions around most probable values, which should be
generally different for unimolecular backward and bimolecular
forward elementary reactions, we would immediately obtain that the
particles' mean concentrations do not obey the LMA and that the
chemical equilibrium state is not a true equilibrium. The
situation becomes even more ambiguous if the elementary reaction
rates $k_{+}(\mu)$ and $k_{-}(\mu)$ are not simple bell-shaped,
but more realistic complicated functions of $\mu$ having minima
and maxima (see, e.g., Refs.\cite{hanggi,moro}). As we have
already remarked in the Introduction, for such $k_{+}(\mu)$ and
$k_{-}(\mu)$ the choice of $R$ and $\lambda$, at which the
reactive boundary condition is imposed, has a large degree of
arbitrariness, but its consequences are grave: when one constrains
himself to "contact approximation" and supposes, as it has been
done in Refs.\cite{cardy} and \cite{gopich3}, that two ill-defined
reaction radii are equal to each other, one finds that there are
no corrections to the LMA as $t \to \infty$ and that the CE is a
true TES. On contrary, letting $R$ and $\lambda$ be different on
an arbitrarily small but fixed value, one gets that Eq.(\ref{act})
is always different from Eq.(\ref{lma}), and that both
$a_{\infty}$ and $b_{\infty}$ are dependent on such "kinetic"
parameter as particles diffusion coefficient $D$. Which, in turn,
implies that the asymptotic $t = \infty$ state is not a true
thermodynamic equilibrium, but rather a non-equilibrium
steady-state.

Consider next the explicit form of corrections to the LMA in Eq.(\ref{act})
for microscopically inhomogeneous contact reactions
in three dimensional systems. Here, we have
for the Fourier transformed reaction constants:
\be
k_{+}(p) = K_{+} \frac{\sin(p R)}{p R}, \;\;\; \text{and} \;\;\; k_{-}(p) = K_{-} \frac{\sin(p
\lambda)}{p \lambda},
\fin
and consequently,
\be
\disp J_{+}(|\mu|=R,R) - \frac{K_{+}}{K_{-}} J_{-}(|\mu|=R,\lambda) =
\frac{K_{+}}{2 \pi^2 R} \int^{\infty}_0 p dp \sin(p R) F(p) \left(\frac{\sin(p R)}{p R} -
\frac{\sin(p \lambda)}{p \lambda}\right).
\fin
Expanding next $F(p)$ in Eq.(\ref{eff}) into elementary fractions and
performing the resulting integrals, we find that
 for sufficiently small $R$ and $\lambda$ ($\lambda>R$),
\begin{equation}
\disp J_{+}(|\mu|=R,R) - \frac{K_{+}}{K_{-}} J_{-}(|\mu|=R,\lambda) \approx \frac{K_+}{4\pi D R}(1-\frac{R}{\lambda})
\end{equation}
Substituting the last equation into Eq.(\ref{act}), we arrive eventually at the result presented
in Eq.(\ref{pp}), which has been previously obtained in Ref.\cite{gleb0} in terms of a suitably extended
heuristic Smoluchowski approach. In the next section we will show using a systematic diagrammatic expansion that, curiously enough,
this equation is \textit{exact} in the linear order in deviation
from the equilibrium.

\section{The asymptotic $t = \infty$ state: Exact results.}

In this section we will  analyze behavior of the pairwise correlation function $\sigma_{\alpha \alpha}(\mu)$ entering Eq.(\ref{eps}) in terms of a systematic diagrammatic expansion. In terms of this
expansion, we will derive, in the linear with respect to the deviation from equilibrium approximation,
an integral equation obeyed by $\sigma_{\alpha \alpha}(\mu)$. We proceed
to show then that our criterion
 of the validity of the LMA and of the thermodynamic equilibrium, obtained in the previous section
within the decoupling procedure, is \textit{exact}. Moreover, we
will define corrections to the LMA in Eq.(\ref{lma}) for
microscopically inhomogeneous reactions with contact and
exponential reaction rates.

We start by rewriting
the Langevin equations (\ref{lan}) in the following matricial form:
\be
M\cdot\left(\begin{array}{l}
\delta\al(x,t)\\
\delta\b(x,t)
\end{array}\right)
= \left(\begin{array}{l}
u_1 -2{\epsilon_0}+\zeta\\
u_2+{\epsilon_0}
\end{array}\right)
\fin
where $\delta\al = \delta\al(x,t)$ and $\delta\b = \delta\b(x,t)$ are
local deviations from the stationary values,
Eq.(\ref{dev}), $\zeta = \zeta(x,t)$, $\epsilon_0\equiv \epsilon(p=0)_\infty$,
the functionals $u_{1,2}$ are given by:
\be
\left\{\begin{array}{l}
u_1(x,t)=\disp-\int d\mu  k_+(\mu)\delta\al(x,t)(\delta\al(x-\mu,t)+\delta\al(x+\mu,t)),\\
u_2(x,t)=\disp\int d\mu  k_+(\mu)\delta\al(x-\mu/2,t)\delta\al(x+\mu/2,t),
\end{array}\right.
\fin while the linear operator $M$ is defined in the
Laplace-Fourier space of parameters $s,p$ as:
\begin{eqnarray} \label{m} \disp M(s,p)&=& \int\; \int d\mu \; dt \;
e^{\disp - st + i (p \cdot \mu) } M(\mu) =\nonumber\\
&=& \left(\begin{array}{ll}
s+Dp^2+2 a_\infty(K_{+}+k_+(p)) & - 2 k_-(p)\\
- 2 a_\infty k_+(p)& s+Dp^2+K_-
\end{array}\right).
\end{eqnarray}
 Note that in Eq.(\ref{m}), we have supposed, as in the
previous section, that $k_{\pm}(\mu)$ depend only on the absolute
value of $\mu$ and hence, that $k_{\pm}(p) = k_{\pm}(-p)$.

Denoting next the propagator $G$ as an inverse of the matrix $M$,
 $G=M^{-1}$,
and taking the inverse Laplace transform, we find that the
Fourier-transformed pair correlation function $\sigma_{\alpha
\alpha}(\mu)$, Eq.(\ref{cor}), obeys:
\begin{eqnarray}
\label{sig}
\sigma_{\alpha \alpha}(p) &= &
\left<\left[G_{\al\al}(p,t)*(u_1(p,t)-2 \epsilon_0 \delta(p)+\zeta(p,t)+
\delta\al_0(p)\delta(t))\right]^2\right> \nonumber\\
&+&\left<\left[G_{\al\b}(p,t)*(u_2(p,t)+\epsilon_0\delta(p)+\delta\b_0(p)\delta(t))\right]^2\right>\\
 & &+2\left<\left[G_{\al\al}(p,t)*(u_1(p,t)-2\epsilon_0\delta(p)+\zeta(p,t)+
\delta\al_0(p)\delta(t))\right]\right.\nonumber\\
 & &\left.\times\left[G_{\al\b}(p,t)*(u_2(p,t)+\epsilon_0\delta(p)+\delta\b_0(p)\delta(t))\right]\right>,
\nonumber
\end{eqnarray}
where the symbol $"*"$ denotes the time convolution, while $\delta\al_0$ and $ \delta\b_0$
stand for the initial values of the
Fourier spectra of fluctuations
$\delta \al(x,0)$ and $\delta \b(x,0)$.
Note, however, that  $\delta\al_0$ and $ \delta\b_0$ give rise to exponentially decreasing terms and
are insignificant for both the
asymptotic state and for the long-time kinetic behavior.
Eq.(\ref{sig}) is complimented by the following equation determining
the noise-noise correlation, Eq.(\ref{noises}),
in time--momentum space:
\be
\label{noise}
\left<\ze(p,t)\ze(-p,t')\right>=\disp -
\delta(t-t')\Big(\epsilon(p)-\Omega(p)\Big)\fin
where $\epsilon$ and  $\Omega$ have been defined in the previous section.

Now, following Refs.\cite{cardy} and \cite{droz}, we set up a systematic
diagrammatic expansion of
the pairwise correlation function $\sigma_{\alpha \alpha}(p)$,
defined by Eqs.(\ref{sig}) and (\ref{noise}). This expansion can be obtained either
 from the analysis of the action of the field theory describing the model \cite{droz}
 or more directly by iterating Eq.(\ref{sig}) and collecting the different terms
 appearing during this procedure.
 The
propagators and vertices corresponding to
such an expansion are defined in figure \ref{Fig1}. The two first kinds of vortices (of coordinance 3) arise because of the coupling of the quadratic term in $u_1$ and $u_2$ either with $\delta\al$ or with $\delta\b$; the next one is due to the noise term $\zeta$ and the last one to the $\epsilon_0$ term.

\begin{figure}[ht]
\begin{center}
\includegraphics*[scale=0.6]{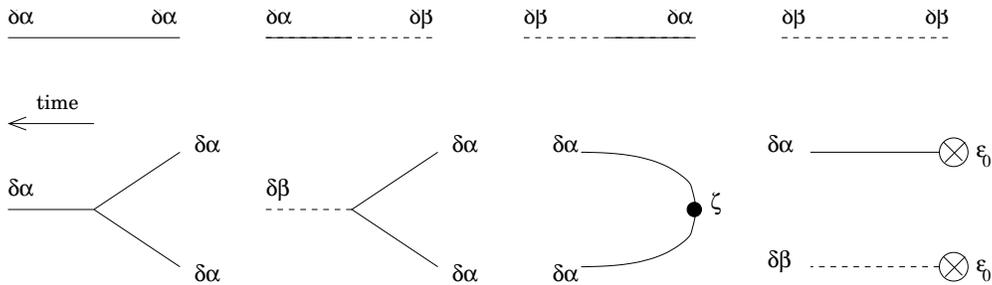}
\caption{\label{Fig1} {Set of propagators (upper part) and
vertices involved in the diagrammatic
 expansion of the correlation function
$\sigma_{\alpha \alpha}(\mu)$.}}
\end{center}
\end{figure}

\begin{figure}[ht]
\begin{center}
\includegraphics*[scale=0.6]{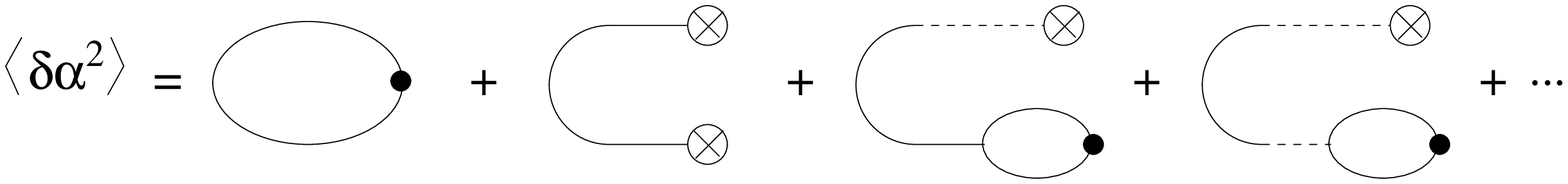}
\caption{\label{Fig2} {Diagrammatic
expansion of the correlation function $\sigma_{\alpha \alpha}(p)$.}}
\end{center}
\end{figure}
A direct summation of these diagrams is impossible, of course. We
will therefore deem this expansion as a perturbative development
around the homogeneous situation, that is, an expansion
 around a configuration
of the elementary reaction rates satisfying $\epsilon(p)_\infty
\equiv 0$. The small parameter in such an expansion will be the
norm (supremum, for instance, denoted as $|\epsilon|$) of the
function $\epsilon(p)_{\infty}$. First order contribution can be
obtained as follows: Note that a vortex $\epsilon_0$ in figure
\ref{Fig1}, corresponding to the term $\epsilon_0$ in
Eq.(\ref{sig}), is of order $|\epsilon|$; likewise, following
Eq.(\ref{noise}), one has that a vortex $\zeta$ is also of order
$|\epsilon|$. The first diagram of figure (\ref{Fig2}) is
therefore of order 1, the second of order 2, and all the following
diagrams are of higher order in $|\epsilon|$. Consequently, in the
linear order in deviation from the equilibrium, we obtain the
following relation: \be \sigma_{\alpha \alpha}(p)=\disp -\int_0^t
dt' G_{\al\al}^0(p,t-t')^2\Big[\epsilon(p)-\Omega(p)\Big],
 \fin
where $G^0_{\al\al}(p,t)$ is the 0--order  propagator, which can be readily obtained in an
explicit form  from Eq.(\ref{m}):
\be\label{0prop}
 G^0_{\al\al}(p,t)=\disp \frac{(K_--q^-)}{q^+-q^-}\disp e^{\disp -(Dp^2+q^-)t}+ \frac{(K_--q^+)}{q^--q^+}\disp e^{\disp -(Dp^2+q^+)t}
\fin where \begin{eqnarray}\label{q+} q^\pm&=&\disp \Big[K_-+2a_\infty \Big(K_++k_+(p)\Big)\pm\sqrt{q}\Big]/2,\\
q&=&\disp \Big[K_-+2a_\infty \Big(K_++k_+(p)\Big)\Big]^2\nonumber\\
&-&8a_\infty\Big(K_-(K_++k_+(p))-2k_+(p)k_-(p)\Big).
\end{eqnarray} Turning next to the limit $t \to \infty$, making use of
Eq.(\ref{eps}) and denoting
 \be {\cal F}^0(p)=\int_0^\infty
dt' G_{\al\al}^0(p,t')^2, \fin
we get
the following system of implicit
equations:
\be\label{implicit}
\left\{\begin{array}{l}
\sigma_{\alpha \alpha}(p)_\infty=\disp \ {\cal F}^0(p)\Big[\Omega(p)_\infty-\epsilon(p)_\infty\Big],\\
\epsilon_0-\Omega(0)_\infty =0,
\end{array}
\right.
\fin
Note now that Eq.(\ref{implicit}) has the same form as Eq.(\ref{imp}),
obtained in section 3 within the framework of the decoupling approximation. The only difference is in the form of the kernel  ${\cal F}^0(p)$, which is  a little bit more complicated than $F(p)$ in Eq.(\ref{imp}) but has exactly the same poles. Note, however, that this difference does not affect the criterium we have formulated in the previous section.
One may immediately deduce from Eq.(\ref{implicit}) that
$\sigma_{\alpha \alpha}(\mu)_\infty \equiv 0$, which implies that the LMA
in Eq.(\ref{lma}) is valid, if and only if $\epsilon(p)$ equals $0$,
which holds only for systems with \textit{microscopically homogeneous} reactions.
This condition can be generalized to all orders and solves exactly
our original problem in a very general framework for reactions in Eq.(\ref{chem}): the chemical
equilibrium is a true thermodynamic equilibrium and the LMA in Eq.(\ref{lma})
 holds {\it only} for \textit{microscopically homogeneous} diffusion-limited  reversible reactions.
 If the reaction is not \textit{microscopically homogeneous}, the asymptotic $t = \infty$ state $(a_\infty,b_\infty)$ in the linear with respect to deviation from equilibrium approximation is described in a closed form by the system in Eqs.(\ref{implicit}).

\subsection{"Contact reactions" approximation revisited.}

In this subsection we turn back to the "contact reaction" case and
re-examine it in within the framework of Eqs.(\ref{implicit}),
which are \textit{exact} in linear with respect to deviation from
equilibrium approximation. As in section 3, we take the reaction
rates in form of delta-functions:  $k_+(\mu)= \gamma_d^{-1}(R)
K_{+} \delta^d(|\mu| - R)$ and $k_-(\mu)=\gamma_d^{-1}(\lambda)
K_- \delta^d(|\mu|-\lambda)$ and suppose that $\lambda>R$.

>From Eq.(\ref{0prop}), we have that  the Fourier-transformed
0--order propagator: \be \label{00} G_{\al\al}^0(p,t)=
\frac{K_-}{4K_+a_\infty+K_-}\disp e^{\disp -Dp^2t}+
\frac{4K_+a_\infty}{4K_+a_\infty+K_-} \disp e^{\disp
-(Dp^2+4K_+a_\infty+K_-)t} \fin Integrating Eq.(\ref{00}) over the
time variable, we get that the kernel ${\cal F}^0(p)$ in
Eq.(\ref{implicit}) is given explicitly by: \be {\cal
F}^0(p)=\disp
\frac{1}{\Big(4K_+a_\infty+K_-\Big)^2}\left(\frac{K_-^2}{2Dp^2}+
\frac{16 K_{+}^2
a_{\infty}^2}{2\Big(Dp^2+4K_+a_\infty+K_-\Big)}+\frac{8 K_+
K_-a_\infty}{2 Dp^2+4K_+a_\infty+K_-}\right) \fin Note that this
function of $p$, in the limit $p \to 0$, has exactly  the same
poles and shows exactly the same asymptotic behavior as $F(p)$,
Eq.(\ref{eff}), obtained using an uncontrollable decoupling
procedure. Solving next Eqs.(\ref{implicit}), we find that in
three-dimensions particles´ mean concentrations $a_{\infty}$ and
$b_{\infty}$ obey \be \label{p} \frac{a_{\infty}^2}{b_{\infty}} =
\frac{K_-}{K_+} \Big[ 1 + \frac{K_+}{4 \pi D R} \left(1 -
\frac{R}{\lambda}\right) + {\cal O}\Big((\lambda - R)^2\Big)\Big]
\fin This relation between the particles´ mean concentrations is
obtained in an exact and controllable way. Quite surprisingly, in
the linear with the respect to the difference $(\lambda - R)$
approximation, it coincide with Eq.(\ref{pp}), obtained in
Ref.\cite{gleb0} within the framework of a suitably extended
\textit{heuristic} Smoluchowski approach.

Similar to Eq.(\ref{p}) results can be also obtained for low-dimensional systems.
Performing straightforward calculations, we find then
that $a_{\infty}$ and $b_{\infty}$ obey:
\be
\frac{a_{\infty}^2}{b_{\infty}} = \frac{K_-}{K_+} \Big[ 1 + \frac{K_+}{4 \pi D R} \left(\lambda-R\right)+ {\cal O}\Big((\lambda - R)^2\Big)\Big],
\fin
and
\be
\frac{a_{\infty}^2}{b_{\infty}} = \frac{K_-}{K_+} \Big[ 1 + \frac{K_+}{4 D } \left(\lambda-R\right)+ {\cal O}\Big((\lambda - R)^2\Big)\Big]
\fin
for two- and one-dimensional systems, respectively.

\subsection{Long-range, exponential reaction probabilities.}

Here we estimate corrections to the LMA for a
 particular example of
long-range distance-dependent rates, characterized by an
exponential dependence on the interparticle separation for the
forward bimolecular reaction, and an exponential dependence on the
radius of a geminate pair born in the elementary act of the
unimolecular backward reaction. We focus on three-dimensional
systems in which $k_{+}(\mu)$ and $k_-(\lambda)$ obey: \be
\label{long} k_{+}(\mu)=\disp \frac{K_+}{8\pi R^3}e^{\disp
-\mu/R}\ \ {\rm and} \ \ k_{-}(\mu)=\disp \frac{K_-}{8\pi
\lambda^3}e^{-\disp \mu/\lambda} \fin For such a choice of
reaction probabilities, we are not in position to solve our
implicit equation (\ref{implicit}) exactly and we have to resort
to an approximate scheme. We thus develop an approximate approach,
supposing that $R$ and $\lambda$ are sufficiently small, such that
the characteristic length $\Lambda_{corr}$ of correlations in
particles´ distributions emerging due to ongoing
\textit{microscopically inhomogeneous} elementary reactions in
Eq.(\ref{long}), is much greater than  $R$ and $\lambda$. This is
precisely the approximation used in section 3 to analyze behavior
of correlations in general case of distance-dependent reactions.
Under such an assumption, we may expect that $\sigma_{\alpha
\alpha}(\mu)_{\infty}$ varies much slower than  $k_{+}(\mu)$ and
$\Omega(p)_{\infty}$ can be represented as in Eq.(\ref{40}). This
implies that \be \label{sigf} \sigma(p)_\infty\approx b_\infty K_-
{\cal
F}^0(p)\left[\frac{1}{(1+p^2\lambda^2)^2}-\frac{1}{(1+p^2R^2)^2}\right]
\fin Evaluating next \be \Omega(0)_\infty\approx
-K_+\int\frac{d^3p}{(2\pi)^3}\sigma(p)_\infty, \fin we get, using
Eq.(\ref{eps}), that particles´ mean concentrations $a_{\infty}$
and $b_{\infty}$ obey \be \frac{a_{\infty}^2}{b_{\infty}} \approx
\frac{K_-}{K_+} \Big[ 1 + \frac{K_+}{16 \pi D R} \left(1 -
\frac{R}{\lambda}\right)\Big], \fin which holds for $R$ and
$\lambda$ sufficiently small. Note that corrections to the LMA in
case of long-range reactions rates in Eq.(\ref{long}) appear to be
smaller than in the "contact" case.

Now, in order to verify that our approximation is
self-consistent, we have to
evaluate the form of pairwise correlation function
$\sigma_{\alpha \alpha}(\mu)$ in $\mu$-space.  Performing the inverse
Fourier transformation of $\sigma_{\alpha \alpha}(p)_{\infty}$ in Eq.(\ref{sigf}),
we find that
for sufficiently small $R$ and $\lambda$ the pairwise correlations follow:
\be \label{pr}
\begin{array}{ll}
\sigma(\mu)_\infty= &\disp  \frac{b_\infty K_-(\lambda^2-R^2)}{4\pi D^2\mu}\left[\frac{16 K^2_{+} a_{\infty}^2}{4 K_{+} a_{\infty} + K_{-}} \disp e^{\disp -\mu/\Lambda_{corr}}+\frac{4 K_{+} K_{-} a_{\infty}}{2 \Big(4 K_{+} a_{\infty} + K_{-}\Big)}\disp e^{\disp -\mu/\sqrt{2} \Lambda_{corr}}\right]\\
   &+ C_R e^{-\disp \mu/R}+ C_\lambda e^{\disp -\mu/\lambda}
\end{array}
\fin where $C_R$ and $C_\lambda$ are $\mu$-independent constant and $\Lambda_{corr}$ has been defined in section 3, Eq.(\ref{cl1}) in terms of the decoupling procedure. Therefore, our assumption underlying the derivation of
the result in Eq.(\ref{pr}) is justified when $\Lambda_{corr} \gg R$, i.e., for sufficiently slow reactions such that the chemical time $\tau_{chem} = (4 K_{+} a_{\infty} + K_{-})^{-1}$ is much larger
than the time necessary to diffuse on distance $R$.
Note also that
the dependence
of the correlation length on the diffusion coefficient is
symptomatic of a non-equilibrium state.

\section{Conclusions.}

To conclude,
in this paper we re-examined two fundamental
concepts of classical chemical kinetics
- the
notion of "Chemical Equilibrium" and the "Law of Mass Action" -
on example of diffusion-limited reversible
$A+A \rightleftharpoons B$ reactions.
We have considered a rather
general lattice
model of such
 reactions, in which the elementary reaction rates are long-ranged and dependent
on the instantaneous distance between particles. This model has
been first analyzed by Zeldovich and Ovchinnikov [18] and pertains
to chemical reactions between excited molecules or reactions
involving transport of a proton or of an electron. It may be also
viewed as a model of elementary reactions taking into account,
albeit in an idealized fashion, different "microscopic" effects
such as solvent structure, different angular orientations of
reactive molecules, energy distributions and etc. In this model
any pair of $A$ particles, which perform standard random walks on
sites of a $d$-dimensional hypercubic lattice may associate
forming a $B$ particle at any moment of time $t$ with the rate
$k_+(\mu)$, where $\mu$ is the instantaneous distance separating
these two particles. In the reverse reaction elementary act, a
mobile $B$ particle may spontaneously dissociate with the rate
$k_-(\lambda)$ giving birth to a geminate pair of $A$s separated
by a distance $\lambda$. In our analysis we supposed that
long-range rate $k_+(\mu)$  of bimolecular forward reaction and
the rate $k_-(\lambda)$, describing the birth of a geminate pair
of $A$ particles born at the distance $\lambda$ apart of each
other within the elementary act of the unimolecular backward
reaction, are arbitrary (integrable) functions of $\mu$ and
$\lambda$. In terms of a formally exact approach based on
Gardiner's Poisson representation method \cite{gardiner}, we have
obtained exact non-linear Langevin equations describing the time
evolution of complex-valued Poisson fields, whose mean values
determine the $A$ and $B$ particles' mean concentrations.
Solutions of these Langevin equations in the asymptotic $t =
\infty$ state were obtained via a certain decoupling approximation
and
 a systematic diagrammatic expansion.

>From these solutions, which coincide in the leading order, we have
deduced a general criterion determining the conditions
 when the classical LMA holds and
when the asymptotic $t = \infty$ state is a true thermodynamic
equilibrium. We have shown that this may only happen when the
distance-dependent elementary reaction rates obey a very
restrictive condition of microscopic homogeneity: the ratio $
k_+(\mu)/k_-(\mu)$ does not depend on $\mu$ for any $\mu$. It
seems that such a condition may be considered as apparently
unrealistic since the \textit{bimolecular} forward and
\textit{unimolecular} backward reactions are supported by
different physical processes of classical and quantum origin. At
present time, no general argument exists that it should be always
the case. In case when $ k_+(\mu)/k_-(\mu)$ is $\mu$-dependent,
i.e. the elementary reactions are \textit{microscopically
inhomogeneous}, it appears that the detailed balance is broken,
the LMA does not hold, particles' concentrations are spatially
correlated and, remarkably, both correlation length and particles´
mean concentrations do depend on such kinetic parameter as the
diffusion coefficient. This implies that the CE is not a true
thermodynamic equilibrium but rather a non-equilibrium
steady-state. Diffusion coefficient $D$-dependent corrections to
the LMA have been calculated explicitly in several particular
cases.

Consequently, for diffusion-limited reversible reactions the
diffusional relaxation of the system is not fast enough to offset
the perturbative effect of ongoing \textit{microscopically
inhomogeneous} elementary reactions (breaking the detailed
balance) even in the asymptotic $t = \infty$ state. This results
in a non--vanishing current which modifies the structure of
 fluctuations and globally changes asymptotic concentrations.
We emphasize that such a non-equilibrium
steady-state emerges in a closed system with conserved overall
concentration of particles and products, without any external
inflow of particles. It might be also worthy to remark that
the diffusion-limited reactions provide thus a nice example of physical
systems in which
an arbitrarily small but finite difference (of classical or quantum origin)
between the microscopic rates $ k_+(\mu)$
and $k_-(\mu)$
entails a fundamental change in the asymptotic behavior of a many particle system.

It is important to remark that our work does not contest any
principle of thermodynamics, and in particular the zeroth
principle. Indeed, in  our model the  system reaches a steady
state, which can be described as the stationary point of a
thermodynamical potential $G$, in agreement with the zeroth
principle. Our work shows that $G$ does depend on dynamical
quantities such as the diffusion coefficient $D$, which implies
that the entropic contribution to $G$ can not be postulated
\textit{a priori}, but has to be defined using kinetic approaches
taking into account dynamical effects. On the other hand, the LMA
itself is not a law of thermodynamics, but rather a law of an
ideal gas, since it relies on the hypothesis of infinitely fast
mixing of the reaction bath. Consequently, there is no surprise
that the classic LMA does not hold for systems with
microscopically inhomogeneous reactions and diffusion as the
limiting transport process.

Finally, we note that the analysis presented in our work may be
extended in several directions. In particular, the question of the
corrections to the LMA may be addressed for other types of
reaction schemes, including, e.g., reactions between excited
molecules, in which case particles possess an intrinsic life-time.
As well, one may expect that the effects observed here will become
more pronounced for photochemical, ionization and electrochemical
reactions \cite{bb}.

\vspace{0.3in}

The authors gratefully acknowledge helpful discussions with
S.F.Burlatsky, S.Bratos, I.V.Gopich, M.Hankel, P.H\"anggi,
A.Lesne, K.Lindenberg, Z.Koza, M.Rausher, K.Seki, I.M.Sokolov and
M.Tachiya. GO thanks the Alexander von Humboldt Foundation for the
financial support via the Bessel Research Award.

\end{document}